\documentclass[12pt]{article}

\usepackage{amsmath,amsfonts,amssymb,color,graphicx,overpic,} 
\usepackage{slashed}
\usepackage{epsfig}
\usepackage{amsmath}
\usepackage{ wasysym }
\usepackage{graphicx}
\usepackage[english]{babel}
\usepackage{graphic x}
\usepackage{relsize}

\begin{document}


\renewcommand{\arraystretch}{2}

\begin{titlepage}
\rightline{\large December 2014}
\vskip 2cm
\centerline{\Large \bf
Diurnal modulation signal from} \vskip 0.3 cm
\centerline{\Large \bf dissipative hidden sector dark matter}

\vskip 2.2cm
\centerline{\large R.Foot\footnote{E-mail address: rfoot@unimelb.edu.au}, S.Vagnozzi\footnote{
E-mail address: vagnozzi@nbi.dk, sunny.vagnozzi@fysik.su.se}}

\vskip 0.7cm
\centerline{\it ARC Centre of Excellence for Particle Physics at the Terascale,}
\centerline{\it School of Physics, University of Melbourne,}
\centerline{\it Victoria 3010 Australia}
\vskip 2cm
\noindent

We consider a simple generic dissipative dark matter model: a hidden sector featuring two dark matter particles charged under an unbroken $U(1)'$ interaction. Previous work has shown that such a model has the potential to explain dark matter phenomena on both large and small scales. In this framework, the dark matter halo in spiral galaxies features nontrivial dynamics, with the halo energy loss due to dissipative interactions balanced by a heat source. Ordinary supernovae can potentially supply this heat provided kinetic mixing interaction exists with strength $\epsilon \sim 10 ^{-9}$. This type of kinetically mixed dark matter can be probed in direct detection experiments. Importantly, this self-interacting dark matter can be captured within the Earth and shield a dark matter detector from the halo wind, giving rise to a diurnal modulation effect. We estimate the size of this effect for detectors located in the Southern hemisphere, and find that the modulation is large ($\gtrsim 10\%$) for a wide range of parameters.

 \end{titlepage}
 
 \newpage

\section{Introduction}

Dark matter might plausibly arise within a hidden sector. That is, a sector of additional particles and forces which couple to ordinary matter predominantly via gravity. An interesting class of such hidden sector dark matter arises when the hidden sector features an unbroken $U(1)'$ gauge symmetry. The associated massless gauge boson, the \textit{dark photon}, mediates self-interactions among the dark matter particles which can also be dissipative.

Dissipative dark matter has been studied in the context of mirror dark matter \cite{volkasspheroidal} (MDM, for an up-to-date review see \cite{review}), where the hidden sector is exactly isomorphic to the Standard Model \cite{footlewvolkas}, and more generally in \cite{dhsdm}. In the latter case, focused on here, dark matter consists of two hidden sector particles, $F_1$ and $F_2$, both charged under an unbroken $U(1)'$ symmetry. Within this picture, the halos around spiral galaxies such as the Milky Way are (currently) mainly in the form of a pressure-supported plasma of $F_1$ and $F_2$ particles. The dark matter halo is assumed to have evolved into a steady state configuration where it is in hydrostatic equilibrium and the energy it loses to dissipative interactions (via e.g. thermal dark bremsstrahlung) is balanced by a heat source. Such a dynamically evolved halo appears capable of explaining small-scale structure observations: the inferred cored profile of dark matter halos, the Tully-Fisher relation, and so forth \cite{dhsdm}.

A possible heat source arises if the kinetic mixing interaction exists:
\begin{eqnarray}
{\cal L} _{\text{km}} = \frac{\epsilon}{2}F ^{\mu \nu}F ^{'} _{\mu \nu} \ .
\end{eqnarray}
This interaction endows the dark particles, $F_1$ and $F_2$, with a tiny ordinary electric charge. The studies \cite{review,dhsdm} have shown that kinetic mixing induced processes in the core of ordinary core-collapse supernovae can supply the energy needs of such a halo, if $\epsilon \sim 10 ^{-9}$. That is, these processes are able to generate enough energy (transported to the halo via dark photons) to compensate for the energy lost due to dissipative interactions. This mechanism, and other astrophysical and cosmological considerations, also constrain the masses of the $F_1$ and $F_2$ particles, with one of them being light, $m _{F_1} \sim \rm MeV$, and the other heavier, $m _{F_2} \sim \rm GeV-TeV$.\footnote{Here and throughout the article, natural units with $\hbar = c = k _B = 1$ will be used.} Kinematic considerations then indicate that the processes $F_1$-electron and $F_2$-nuclei scattering will be of particular importance in the context of direct detection experiments.

MeV scale dark matter particles scattering off electrons have been proposed as a mechanism to potentially explain the DAMA \cite{dama} and CoGeNT \cite{cogent} annual modulation signals \cite{electron}. Scattering of the light $F_1$ particles off electrons might thereby explain the annual modulation signals observed by DAMA and CoGeNT. However, the details are quite subtle, as the flux of the light $F_1$ particles in the proximity of the Earth is expected to be strongly influenced by dark electromagnetic fields, generated within the Earth by captured $F_1$ and $F_2$ dark matter. Further details of $F_1$-electron scattering will be postponed to future work. Here we focus on the $F_2$-nuclei scattering detection channel, and study possible diurnal modulation signatures expected due to the effect of captured $F_2$ dark matter which can block the halo wind.
 
\section{Two-component dissipative hidden sector dark matter}

The model we consider comprises a hidden sector consisting of dark matter particles charged under an unbroken $U(1)'$ symmetry, and possibly other interactions which we shall not be concerned with. The dynamics of our theory are described by the Lagrangian:
\begin{eqnarray}
{\cal L} = {\cal L} _{\text{SM}} + {\cal L} _{\text{HS}} + {\cal L} _{\text{mix}} \ ,
\end{eqnarray}
where ${\cal L} _{\text{SM}}$ is the Standard Model Lagrangian, ${\cal L} _{\text{HS}}$ is the hidden sector Lagrangian, and ${\cal L} _{\text{mix}}$ encompasses interaction terms which interconnect the two sectors. The interactions associated with the unbroken $U(1)'$ symmetry (dark electromagnetism) are mediated by a massless gauge boson, the \textit{dark photon} ($\gamma _{_D}$). A particularly simple instance occurs when the hidden sector consists of two Dirac fermions, described by the fields $F_1$ and $F_2$, with masses $m _{F_1}$ and $m _{F_2}$ and dark charges $Q _{F_1} ^{'}$ and $Q _{F_2} ^{'}$, opposite in sign but not necessarily equal in magnitude:\footnote{Replacing the $F_1$, $F_2$ particles with two scalar fields leads to an equally simple model. Furthermore, the diurnal modulation signal to be discussed in the present paper depends in no essential way on the spin of the dark matter particles. For concreteness, though, we here focus on the fermionic model.}
\begin{eqnarray}
{\cal L} _{\text{HS}} = -\frac{1}{4}F ^{'\mu \nu}F ^{'} _{\mu \nu} + \overline{F}_1(iD _{\mu}\gamma ^{\mu} - m _{F_1})F_1 + \overline{F}_2(iD _{\mu}\gamma ^{\mu} - m _{F_2})F_2 \ .
\end{eqnarray}
Here, $F ^{'} _{\mu \nu} = \partial _{\mu}A ^{'} _{\nu} - \partial _{\nu}A ^{'} _{\mu}$ is the field-strength tensor associated with the $U(1)'$ interaction, with $A ^{'} _{\mu}$ being the relevant gauge field. The covariant derivative relevant to this interaction acts on the fermionic fields as $D _{\mu}F _j = \partial _{\mu}F _j + ig'Q ^{'} _{F_j}A ^{'} _{\mu}F _j$, where $g'$ is the coupling constant for the dark electromagnetic interaction. The presence of an accidental $U(1)$ global symmetry, together with the gauge symmetry, implies conservation of $F_1$ and $F_2$ number, and hence stability of the two dark fermions. The particle content of the hidden sector is thus massive, dark and stable, essential characteristics of a suitable dark matter candidate.

In the early Universe, a primordial particle-antiparticle
asymmetry is presumed to set the relic abundance of the
$F_1$ and $F_2$ particles. Any symmetric component is expected to be efficiently annihilated by the dark electromagnetic interactions. This means that the dark matter content of the Universe today is dominated by particles with a potentially negligible amount of anti-$F_1$ and anti-$F_2$ particles.\footnote{Of course, the exact ratio of dark matter antiparticles to particles today depends on the thermal history of the dark and visible sectors in the early Universe. If $T _{\gamma _{_D}} \ll T _{\gamma}$ when $T _{\gamma _{_D}} \approx m _{F_2}$, then the relic abundance of dark antiparticles can be negligibly small for all of the parameter space of interest.  On the other hand if $T _{\gamma _{_D}} \simeq T _{\gamma}$ when $T _{\gamma _{_D}} \approx m _{F_2}$, an upper bound on $m _{F_2}$ of order 8 TeV, 350 GeV, 35 GeV, 3 GeV for $\alpha ' = 10 ^{-1}, 10 ^{-2}, 10 ^{-3}, 10 ^{-4}$ respectively can be derived by requiring the symmetric component to be efficiently annihilated away \cite{petrakinew} (see also \cite{vecchi,vonharling}).} The model is then an example of \textit{asymmetric dark matter}, extensively discussed in the recent literature (see e.g. \cite{asymmetric} and references therein). Dark matter asymmetry and local neutrality of the Universe imply:
\begin{eqnarray}
n _{F_1}Q ^{'} _{F_1} + n _{F_2}Q ^{'} _{F_2} = 0 \ ,
\label{asymmetry}
\end{eqnarray}
where $n _{F_j}$ denotes the $F_j$ particle number density.

The possible interactions described by ${\cal L} _{\text{mix}}$ are strongly constrained by the requirements of gauge invariance and renormalizability. For our model, this restricts ${\cal L} _{\text{mix}}$ to only a kinetic mixing term \cite{foothe}, which leads to photon-dark photon kinetic mixing:
\begin{eqnarray}
{\cal L} _{\text{mix}} = \frac{\epsilon '}{2}F ^{\mu \nu}F ^{'} _{\mu \nu} \ .
\end{eqnarray}
A non-orthogonal transformation can remove the kinetic mixing. The net effect of this interaction is to provide the dark fermions with a tiny ordinary electric charge \cite{holdom}. As a result, the dark fermions couple to the visible photon with charge:
\begin{eqnarray}
g'Q ^{'} _{F_j}\epsilon ' \equiv \epsilon _{F_j}e \ .
\end{eqnarray}
The interactions of $F_1$ with the dark photon are characterized by the dark fine structure constant, $\alpha ' \equiv (g'Q _{F_1})^2/4\pi$, while the coupling of $F_2$ with the dark photon is modified by the charge ratio, $Z' \equiv Q ^{'} _{F_2}/Q ^{'} _{F_1}$. The fundamental physics of this model is described by five parameters: $m _{F_1}$, $m _{F_2}$, $\alpha '$, $Z'$ and $\epsilon \equiv \epsilon _{F_1}$.

The model described above has been thoroughly analysed in the context of early Universe cosmology and galactic structure in \cite{dhsdm}. Its dark matter phenomenology is similar to, but generalizes the MDM case, and is more distantly related to a number of other hidden sector models which feature an unbroken $U(1)'$ interaction (see e.g. \cite{unbroken}). Within the scenario being considered, the dark matter halo in spiral galaxies is presumed to be (currently) in the form of a roughly spherical plasma composed of $F_1$ and $F_2$ particles. The plasma can cool via dissipative processes, for instance thermal dark bremsstrahlung, thus requiring a heat source which can replace this energy lost. It has been argued that kinetic mixing induced processes within the core of ordinary core-collapse supernovae can provide such a heat source, provided $\epsilon \sim 10 ^{-9}$ and $m _{F_1} \lesssim 100 \ {\rm MeV}$. 

The analysis of early Universe phenomenology (including bounds on the number of relativistic degrees of freedom encoded by 
$N _{\text{eff}}[\text{CMB}]$ and $N_{\text{eff}}[\text{BBN}]$) and galactic structure arguments constrained the five parameters of the model \cite{dhsdm}. These considerations indicated a favored region of parameter space for the masses of the two fermions: the lighter particle ($F_1$) with mass in the MeV range and the heavier one ($F_2$) with mass in the GeV-TeV range. Some implications for direct detection experiments of this same model have been considered in \cite{hiddensector}, which also focused on the case $m _{F_1} \ll m _{F_2}$, additionally assuming $|Z'| \gg 1$.
\footnote{An interesting question is whether this model is consistent with measurements on cluster scales, e.g. those associated with the Bullet Cluster. 
The main difficulty in 
addressing this point is that the dark matter distribution on cluster scales is poorly constrained and also very difficult to model. 
Adopting the NFW distribution (or similar) for the cluster dark matter derived from simulations of collisionless dark matter may be unreliable, especially when
self interaction cross sections are large ($\sigma/M \gtrsim 1 \ \rm{cm}^2/\rm{g}$).  
A significant fraction of the dark matter could be bound into galactic or subgalactic-sized halos, or into more compact systems 
such as hypothetical ``dark stars" (see \cite{silagadze,review,dhsdm} for relevant discussions). 
If dark matter is sufficiently clumpy then the dark matter 
associated with each cluster would pass through each other essentially unimpeded, potentially consistent with the observations \cite{bullet}.}  

In the dark halo of the Milky Way,
the dark electromagnetic interactions are expected to keep the particles in thermal equilibrium, at a common temperature $T$. In the proximity of the Earth, under the assumption of hydrostatic equilibrium, this temperature can be roughly estimated \cite{hiddensector}:
\begin{eqnarray}
T \simeq \frac{1}{2}\overline{m}v _{\text{rot}} ^2 \ .
\end{eqnarray}
Here, $v _{\text{rot}} \approx 220 \ \rm km/s$ is the Milky Way's rotational velocity, while $\overline{m}$ designates the mean mass of the particles in the dark plasma which, in the two-component case we are considering, is given by:
\begin{eqnarray}
\overline{m} = \frac{n _{F_1}m _{F_1} + n _{F_2}m _{F_2}}{n _{F_1} + n _{F_2}} \ .
\label{meanmass}
\end{eqnarray}
The galaxy structure arguments of \cite{dhsdm} indicate that the Milky Way halo could be nearly fully ionized, except for the K-shell atomic states, so that $|Z'| \geq 3$ (see \cite{dhsdm} for further details). It follows that the mean mass can be approximated as being $\overline{m} \approx m _{F_2}/(|Z'|-1)$.

In a reference frame with no bulk halo motion, we expect the distribution function of the halo dark matter particles to be Maxwellian. The velocity dispersion of the $i$-th particle species, $v _0[F _i]$ is mass-dependent, and is given by (see e.g. \cite{hiddensector}):
\begin{eqnarray}
v _0[F _i] \simeq v _{\text{rot}}\sqrt{\frac{\overline{m}}{m _i}} \ .
\label{dispersion}
\end{eqnarray}
Combining Eqs.(\ref{asymmetry},\ref{meanmass},\ref{dispersion}), it follows that $v _0[F _2] \ll v _{\text{rot}}$ for $m _{F_1} \ll m _{F_2}$ and $|Z'| \gg 1$. This mass dependent velocity dispersion is a distinctive feature of this type of dark matter.

The $F_1$ and $F_2$ particles can potentially be observed in direct detection experiments. With $m _{F_1} \sim \rm MeV$ and $m _{F_2} \sim \rm GeV-TeV$, $F_1$-electron and $F_2$-nuclei scattering are expected to be of most interest (essentially Rutherford scattering, possible due to the kinetic mixing induced small electric charge). In principle, both of these interaction channels can be searched for. In this article, we focus on the $F_2$-nuclei scattering channel, and consider the diurnal modulation signal which we will show is a characteristic feature of this type of dark matter.

A diurnal modulation in a direct detection experiment can arise if dark matter particles are captured within the Earth and block the halo dark matter wind. Diurnal modulation due to self-interacting dark matter was first studied in the context of MDM \cite{footdiurnal}. A diurnal modulation effect can also ensue following interactions of dark matter particles with the constituent nuclei of the Earth \cite{collar}. This can be important for some models, such as the case of light GeV scale dark matter (and could be important in our case for part of the parameter space) \cite{shoemaker}. Here we focus on the diurnal modulation effect arising from self-interactions between halo and captured dark matter particles, which we show is large for a wide range of parameter space.

\section{Dark matter shielding radius}

A distinctive feature of this model, and of hidden sector models in general, is the self-interacting nature of the dark matter particle content. The self-interactions can lead to a significant quantity of dark matter being captured within the Earth, potentially blocking the $F_2$ dark matter galactic halo wind. In this section we will quantify this effect, by estimating the \textit{shielding radius} due to dark matter capture.

Initially, $F_2$ particles will occasionally be captured by the Earth, through hard scattering processes of $F_2$ on constituent nuclei within the Earth, and thus accumulate inside our planet (cf. \cite{foot94}). When a sufficient number have accumulated, $F_2$ particles will be captured following self-interactions. Let us define $d _{\min}$ to be the distance of closest approach to the center of the Earth of a halo $F_2$ particle, for a given trajectory. The shielding radius, $R_s$, is the maximum value $d _{\min}$ can take for which the incoming $F_2$ particle will be captured due to self-interactions with the Earth bound dark matter. That is, halo $F_2$ particles with trajectories having $d _{\min}<R _s$ will be captured and accumulate within the Earth.
This means that $F_2$ particles will be captured at the ``geometric" rate given by:
\begin{eqnarray}
\frac{d{\cal N}}{dt} \approx \pi R _s ^2v _{\text{rot}}n _{F_2} \ ,
\end{eqnarray}
where $v _{\text{rot}} \simeq 220 \ \rm km/s$ is the galactic rotational velocity, $n _{F_2}$ is the number density of halo $F_2$ particles ($n _{F_2} = \rho _{\text{dm}}/m _{F_2}$, with $\rho _{\text{dm}} \approx 0.3 \ \rm GeV/cm ^3$ at the Earth's location). Here ${\cal N}$ denotes the total number of captured $F_2$ particles accumulated within the Earth.

In the analysis to follow, we assume no significant initial population of dark matter particles in the Earth. In fact, during the formation of the Solar System, we expect dark matter particles to be captured within the newly forming Earth. To estimate the ``initial" number of captured dark matter particles would require us to model the formation of the Solar System, and is beyond the scope of this paper. Here we simply note that, by assuming that a negligible number of dark matter particles are captured initially, the derived shielding radius will be underestimated. It follows that the diurnal modulation signal can potentially be maximal for a larger range of parameters than those given.

To gain further insight into the process of dark matter capture, and hence estimate the shielding radius, we have to determine the density distribution of captured $F_2$ particles within the Earth, $N _{F_2}(r)$ [spherical symmetry is assumed]. This can be determined from the hydrostatic equilibrium condition, but first we need to work out the temperature profile, $T(r)$.

In addition to $F_2$ particles, the light $F_1$ particles will also be captured in the Earth. The rate of $F_1$ capture is expected to be influenced by dark electromagnetic fields in
such a way as to keep the net $U(1)'$ charge of the Earth small (cf. \cite{electron}). Kinetic mixing induced interactions allow the captured $F_1$ and $F_2$ particles to interact with ordinary nuclei and electrons via Rutherford scattering, with cross-section $d\sigma/d\Omega \propto \epsilon ^2/v ^4$, where $v$ is the relative velocity of the captured $F_1$/$F_2$ particle. Given that the captured particles lose energy rapidly, and hence decrease their velocity, the dependence of the Rutherford scattering cross-section on velocity ($\propto 1/v ^4$) suggests that the $F_1$/$F_2$ particles and ordinary matter will quickly thermalize. That is, the dark matter particles and ordinary matter in the Earth will share a common temperature profile, $T(r)$ [a possible exception is near the ``surface" of the dark matter distribution, where halo heating can be important]. At the relevant temperature range within the Earth, the astrophysical and cosmological constraints derived in \cite{dhsdm} indicate that the $F_1$ and $F_2$ states will combine to form atoms for essentially all of the fundamental parameter space of interest.\footnote{Astrophysical and cosmological considerations were exploited to determine bounds on the kinetic mixing parameter, $\epsilon$, in \cite{dhsdm}. For the heating mechanism arising from ordinary core-collapse supernovae to work, $\epsilon \gtrsim 10 ^{-10}$ is required. Large-scale structure considerations (ensuring that dark acoustic oscillations do not modify LSS early growth) were used to set an upper bound on $\epsilon$: $\epsilon \lesssim 10 ^{-8}(\alpha '/\alpha) ^4(m _{F_1}/{\rm MeV}) ^2({\cal M}/m _e) ^{\frac{1}{2}}$, where ${\cal M} \equiv \max(m _e,m _{F_1})$. These bounds constrain the binding energy of the ``valence" $F_1$ particle, $I \sim {\alpha '} ^2m _{F_1}/2 \gtrsim {\rm eV}$, which is greater than the relevant temperature for essentially all of the fundamental parameter space of interest.} Thus the captured $F_1$ and $F_2$ particles ultimately form a gas of $F_2$ atoms with known temperature profile $T(r)$.

The density $N _{F_2}(r)$ is dictated by gravity and pressure through the hydrostatic equilibrium condition:
\begin{eqnarray}
\frac{dP(r)}{dr} = -\rho (r)g(r) \ .
\label{hydrostatic}
\end{eqnarray}
In the above equation [Eq.(\ref{hydrostatic})], $P(r) = N _{F_2}(r)T(r)$ and $\rho (r) = m _{F_2}N _{F_2}(r)$ are the pressure and mass density profiles of the captured $F_2$ atoms, and $g(r)$ is the local gravitational acceleration:
\begin{eqnarray}
g(r) = \frac{G}{r ^2}\int _0 ^r 4\pi {r'} ^2\rho _E (r') dr' \ .
\end{eqnarray}
Here $\rho _E(r)$ denotes the Earth mass density profile. The hydrostatic equilibrium condition [Eq.(\ref{hydrostatic})] can be rearranged to the form:
\begin{eqnarray}
\frac{dN _{F_2}(r)}{dr} = -\frac{N _{F_2}(r)}{T(r)}\left ( m _{F_2}g(r) + \frac{dT(r)}{dr} \right ) \ .
\label{hydrostatic1}
\end{eqnarray}
Solving Eq.(\ref{hydrostatic1}) entails specifying a form for the Earth temperature and density profiles, $T(r)$ and $\rho _E (r)$ [the latter entering
Eq.(\ref{hydrostatic1}) through the local gravitational acceleration profile, $g(r)$]. Following \cite{footdiurnal}, we adopt a linear approximation for the
profiles obtained from the Preliminary Reference Earth Model \cite{preliminary}. Eq.(\ref{hydrostatic1}) can now be solved to obtain the number density of
captured $F_2$ particles (some examples are shown in Figure 1). Note that the obtained number density profile depends only on the mass of the dark matter particle, $m _{F_2}$, and is independent of the other fundamental parameters.

Having estimated the number density profile of captured $F_2$ particles, it is now straightforward to work out the shielding radius, $R_s$. 
We approximate the trajectories of the incoming dark matter particles by straight lines. Along the trajectory, the distance is traced by 
the coordinate $q$, in such a way that the point of closest approach to the center of the Earth ($r = d _{\min}$) has $q=0$. 
An incoming $F_2$ particle is captured if it loses its energy to self-interactions. We find, following a calculation completely analogous to that in \cite{footdiurnal}, that an $F_2$ particle is captured within the Earth if the following condition is satisfied:
\begin{eqnarray}
\int _{q _{\min}} ^{q _{\max}} n _{F_2}\left ( r = \sqrt{d _{\min} ^2 + q ^2} \right ) dq \gtrsim \frac{E _i ^2}{4\pi{Z'} ^4{\alpha '} ^2\ln \left [ \left (\frac{m _{F_2}}{m _{F_1}} \right ) \left ( \frac{v _{\text{rot}}}{\alpha '} \right ) \right ]} \ ,
\label{capture}
\end{eqnarray}
where, as a measure of the average initial energy of the $F_2$ particles, we take $\langle E _i \rangle \approx m _{F_2}v _{\text{rot}} ^2/2$. In the above, $q _{\max,\min} = \pm \sqrt{R _E ^2 - d _{\min} ^2}$ and $R _E \simeq 6371 \ \rm km$ is the Earth's radius. For a given point in parameter space, 
we can solve Eq.(\ref{capture}) numerically by iterating over increasing values 
\vskip 0.5cm
\centerline{\epsfig{file=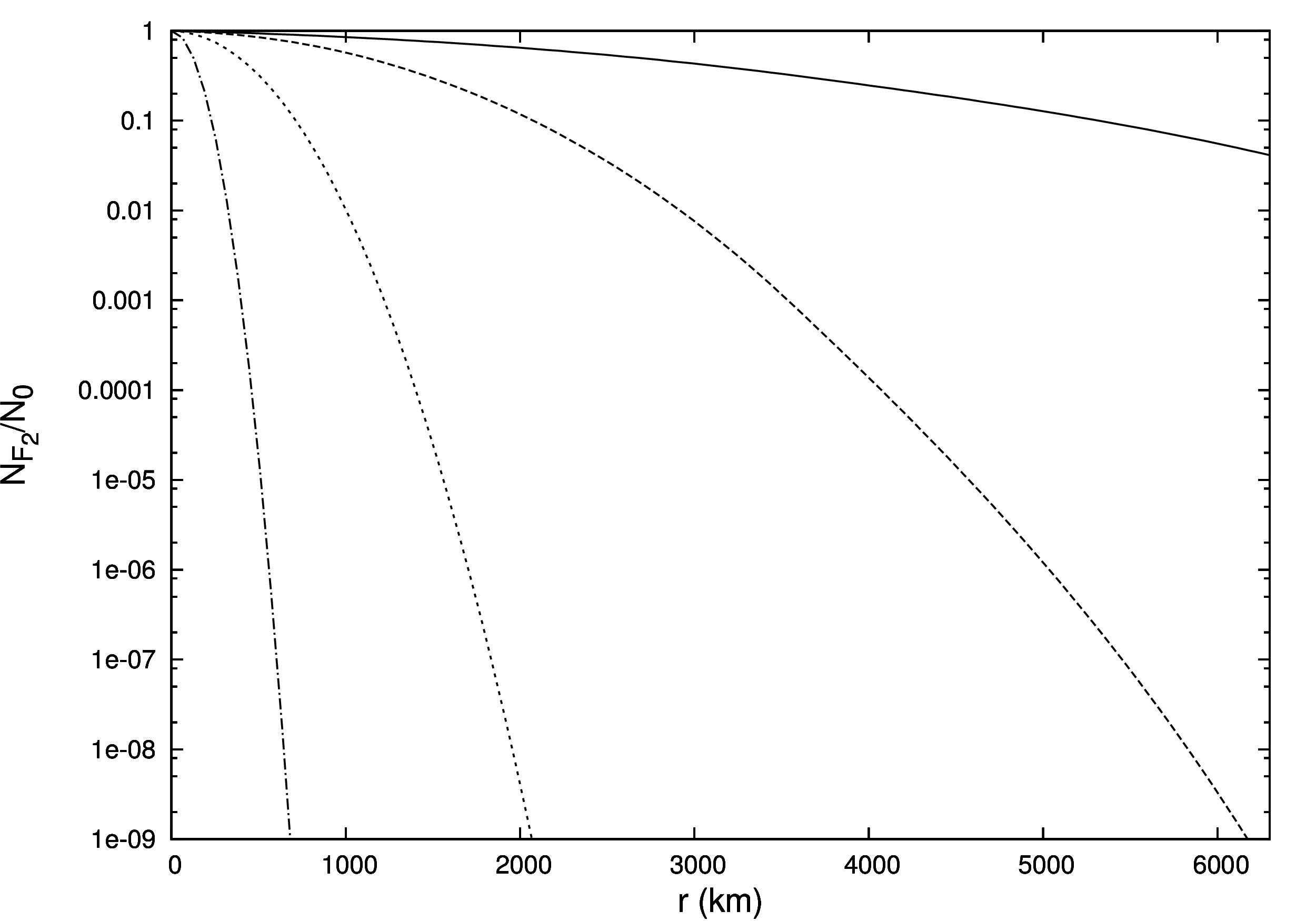, angle=270,width=14.5cm}}
\vskip -0.9cm
\noindent
{\small
Figure 1: Number density profile of captured $F_2$ particles, $N _{F _2} (r)$, normalized to the central number density [$N _0 = N _{F _2} (r=0)$]. 
Curves from right to left: $m _{F_2} = 1 \ {\rm GeV}, 10 \ {\rm GeV}, 100 \ {\rm GeV}, 1 \ {\rm TeV}$.}
\vskip 0.4cm
\noindent
of $d _{\min}$ 
and determining the largest value of $d _{\min}$ for which the left-hand side of Eq.(\ref{capture}) exceeds the 
right-hand side. 
This value defines the shielding radius, $R _s$. Our numerical study determined that the solution displays a very minor dependence on $m _{F_1}$, depending mainly on the remaining three parameters ($m _{F_2}$, $\alpha '$ and $Z'$) and is currently:
\begin{eqnarray}
R _s \simeq 5300 \left ( \frac{\alpha '}{10 ^{-3}} \right ) ^{0.06}\left ( \frac{m _{F_2}}{10 \ \rm GeV} \right ) ^{-0.55}\left ( \frac{|Z'|}{10} \right ) ^{0.14} {\rm km} \ .
\label{shieldingradius}
\end{eqnarray}
The above estimate [Eq.(\ref{shieldingradius})] is valid to a good approximation within the range of parameter 
space: $5 \times 10 ^{-4} \lesssim \alpha ' \lesssim 5 \times 10 ^{-2}$, $5 \ {\rm GeV} \lesssim m _{F_2} \lesssim 300 \ {\rm GeV}$, $3 \lesssim |Z'| \lesssim 40$.

We point out a few caveats. From Figure 1 it can be inferred that $F_2$ particles might be able to escape the Earth if $m _{F_2} \lesssim 5 \ \rm GeV$. Thus, the analysis to follow is strictly only valid for $m _{F_2} \gtrsim 5 \ {\rm GeV}$. Also, for $m _{F_2}$ sufficiently light, the shielding radius can exceed the Earth's radius. If this occurs, our analysis will be invalid and for $R_s$ sufficiently large the halo dark matter wind will be shielded from all directions. This would suppress any diurnal modulation signal. In our analysis we assume that downward going $F_2$ particles are unshielded ($R _s \lesssim R _E$), which from Eq.(\ref{shieldingradius}) implies:
\begin{eqnarray}
m _{F_2} \gtrsim 7 \left ( \frac{\alpha '}{10 ^{-3}} \right ) ^{0.11}\left ( \frac{|Z'|}{10} \right ) ^{0.25} \ \rm GeV \ .
\end{eqnarray}
\vskip 0.5 cm

\section{Diurnal modulation signal}

The captured dark matter particles will shield a dark matter detector located on the Earth from part of the halo dark matter wind. This effect can suppress the rate of $F_2$-nuclei interactions observed in direct detection experiments. Importantly, relative to a given detector location, the direction of the halo wind changes during the day as the Earth rotates. Thus, the amount of shielding of the halo dark matter wind varies during the day, giving rise to a diurnal modulation effect. As we will discuss, given the direction of the Earth's motion through the galaxy, this diurnal modulation effect is expected to be particularly enhanced for direct detection experiments located in the Southern hemisphere.

Let us denote by $\theta _l$ the detector's latitude, by $T _d \simeq 23.9345$ hrs the sidereal day, and by $\theta _h$ the angle subtended by the Earth's motion through the halo with respect to the Earth's spin axis.\footnote{Because of the Earth's motion around the Sun, $\theta _h$ varies slightly during the course of the year. We expect this to give rise to an additional annual modulation of the diurnal modulation effect; here we will ignore this effect, and will simply take the average value $\langle \theta _h \rangle \simeq 43 ^{\circ}$.} We finally denote by $\psi$ the angle between the direction of the Earth's motion through the dark matter halo and the normal vector to the Earth's surface at the relevant detector location. A value $\psi = 0 ^{\circ}$ indicates that the dark matter halo wind is coming vertically down on the detector, while $\psi = 180 ^{\circ}$ indicates that the halo wind is approaching from the other side of the Earth, and hence transiting in proximity of the center of the Earth. Because of the Earth's rotation around its axis, $\psi$ varies during the course of a sidereal day:
\begin{eqnarray}
\cos \psi (t) = \cos \theta _l\sin \left ( 2\pi \frac{t}{T _d} \right )\sin \langle \theta _h \rangle \pm \sin \theta _l\cos \langle \theta _h \rangle \ .
\label{psi}
\end{eqnarray}
In Eq.(\ref{psi}), the $+ [-]$ sign holds for a detector located in the Northern [Southern] hemisphere respectively. This difference in sign plays a crucial
role in the discussion to come, since it implies that $\psi$ can be as large as $\sim 180 ^{\circ}$ only in the Southern hemisphere, as shown in Figure 2. Hence, a diurnal modulation signal is expected to be much more pronounced in the Southern hemisphere since, for part of the day, the dark matter particles are unable to reach the detector, having been blocked by the captured dark matter particles within the Earth.

\vskip 0.8 cm
\centerline{\epsfig{file=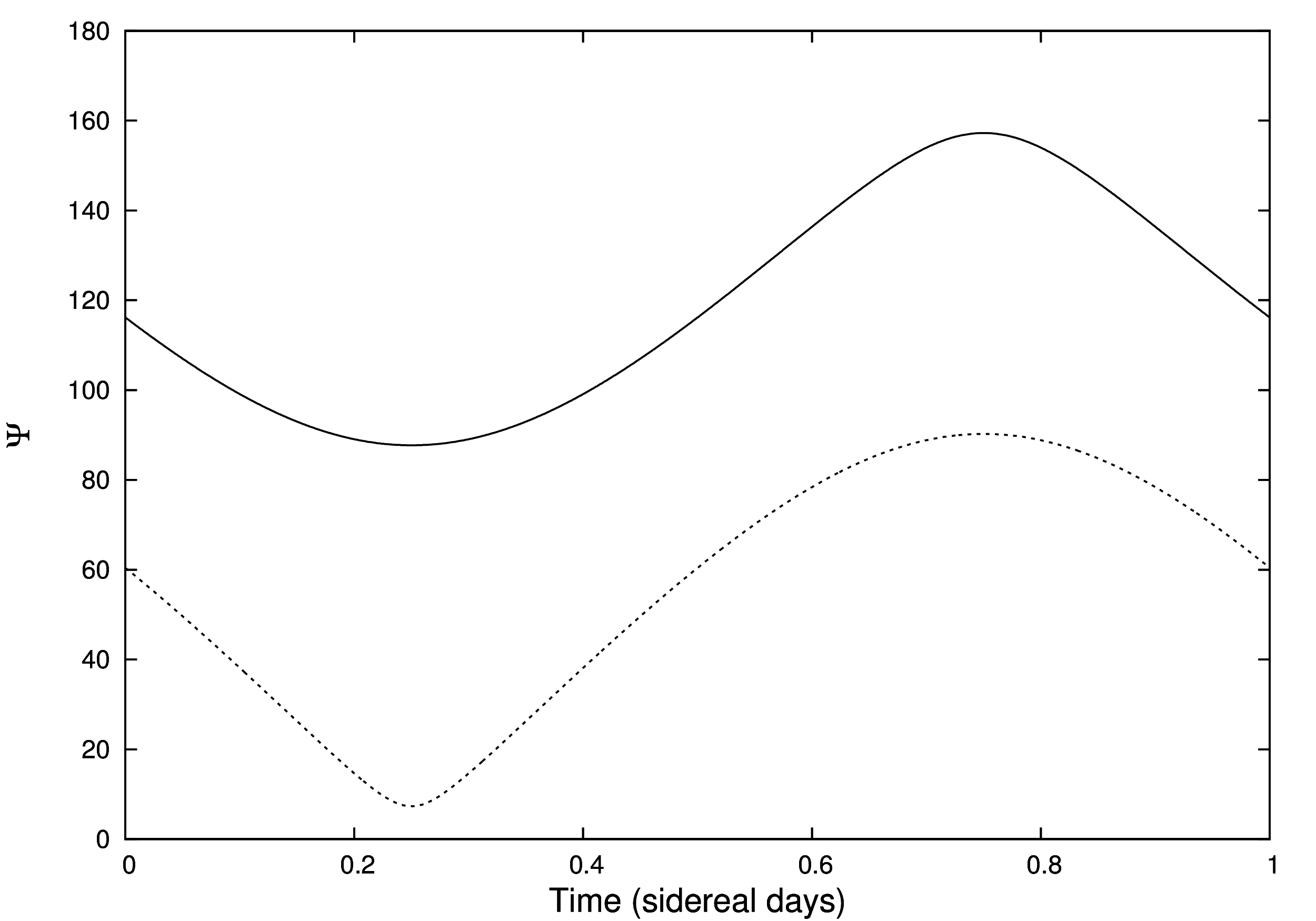, angle=270, width=14.5cm}}
\vskip -0.9cm
\noindent
{\small
Figure 2: Variation of $\psi (t)$ during the course of a sidereal day for a detector located in the 
Stawell mine (solid line) and under the Gran Sasso d'Italia (dashed line).}
%
\vskip 0.4 cm

We now proceed to quantify this suppression of the interaction rate due to dark matter capture. Let us define $\mathbf{v}$ to be the velocity of the halo dark matter particles relative to the Earth (with $v = |\mathbf{v}|$ being the magnitude of this velocity). Additionally define $\mathbf{v}_E$ to be the velocity of the Earth relative to the galactic halo ($\langle |\mathbf{v} _E| \rangle \simeq 220 \ \rm km/s$). In the absence of any shielding of the halo dark matter particles, the differential interaction rate of $F_2$ scattering off target nuclei is given by (see e.g. \cite{footdiurnal}):
\begin{eqnarray}
\frac{dR}{dE _R} = \frac{N _Tn _{F_2}}{(\pi v _0 ^2) ^{\frac{3}{2}}}\int _{|\mathbf{v}| > v _{\min}(E _R)} ^{\infty} \frac{d\sigma}{dE _R}ve ^{-\frac{(\mathbf{v} + \mathbf{v} _E) ^2}{v _0 ^2}} d^3v \ ,
\label{drder}
\end{eqnarray}
where $N _T$ denotes the number of target atoms per kg of detector, $n _{F_2}$ is the number density of halo $F_2$ particles (not to be confused with $N _{F_2}(r)$, the number density of captured $F_2$ particles), and $d\sigma /dE _R$ is the relevant interaction cross-section [$d\sigma/dE _R = F_T ^22\pi \epsilon ^2{Z'} ^2Z ^2\alpha ^2/(m _TE _R ^2v ^2)$, where $m_T$ is the mass of the target nuclei and $F_T$ is the form factor which accounts for their finite size]. Additionally, $v _0$ is defined by Eq.(\ref{dispersion}), $v _{\min} = \sqrt{(m _T + m _{F_2}) ^2E _R/(2m _Tm _{F_2} ^2)}$ is a lower velocity limit determined by kinematics, and $E_R$ is the relevant recoil energy at which we wish to investigate the modulation effect. 

Barring constant factors, and accounting for the dependence of the interaction cross-section on $v$, the interaction rate of halo $F_2$ particles with the target nuclei in the detector is proportional to the quantity:
\begin{eqnarray}
{\cal I} _0 \equiv \int _0 ^{2\pi} d\phi \int _{-1} ^{1} d(\cos \theta) \int _{v _{\min}} ^{\infty} ve ^{-\frac{(\mathbf{v} + \mathbf{v} _E) ^2}{v _0 ^2}} dv \ .
\label{cali}
\end{eqnarray}
To account for dark matter capture and hence shielding of the dark matter halo wind, we multiply the integrand of Eq.(\ref{cali}) by a Heaviside step function:
\begin{eqnarray}
{\cal I}[\psi (t)] \equiv \int _0 ^{2\pi} d\phi \int _{-1} ^{1} d(\cos \theta) \int _{v _{\min}} ^{\infty} ve ^{-\frac{(\mathbf{v} + \mathbf{v} _E) ^2}{v _0 ^2}} H[d _{\min}(\theta , \phi , \psi) - R _s] dv \ .
\label{ipsit}
\end{eqnarray}
Recall $R _s$ is the shielding radius, given in Eq.(\ref{shieldingradius}). The distance of closest approach, $d _{\min}$, is given by \cite{footdiurnal}:
\begin{eqnarray}
d _{\min} = \begin{cases}
            R _E\sqrt{1 - g ^2(\theta , \phi , \psi)} \ , \ & \text{if} \ g(\theta, \phi , \psi) \geq 0 \ , \\
            R _E \ , \ & \text{if} \ g(\theta , \phi , \psi) < 0 \ ,
            \end{cases}
\end{eqnarray}
where $g(\theta , \phi , \psi) \equiv \sin \theta \sin \phi \sin \psi - \cos \theta \cos \psi$.

We can now evaluate ${\cal R}(t)$, the percentage rate suppression due to dark matter shielding, where ${\cal R} = 100 \%$ indicates a total suppression of the interaction rate:
\begin{eqnarray}
{\cal R}(t) = 100\left (1 - \frac{{\cal I}[\psi (t)]}{{\cal I} _0} \right ) \% \ .
\label{calr}
\end{eqnarray}
In Figures 3,4 we present results for ${\cal R}(t)$ for proposed detectors located in the Stawell mine (near Melbourne, $\theta _l \simeq 37.1 ^{\circ}$) and in the Andes Lab (on the Argentinean-Chilean border, $\theta _l \simeq 30.2 ^{\circ}$). As the figures show, the diurnal modulation effect can be very large for these Southern hemisphere detectors. For a detector located in the Northern hemisphere, such an effect is instead expected to be much smaller, and hence more difficult to observe.


%
\vskip 0.5cm
\centerline{\epsfig{file=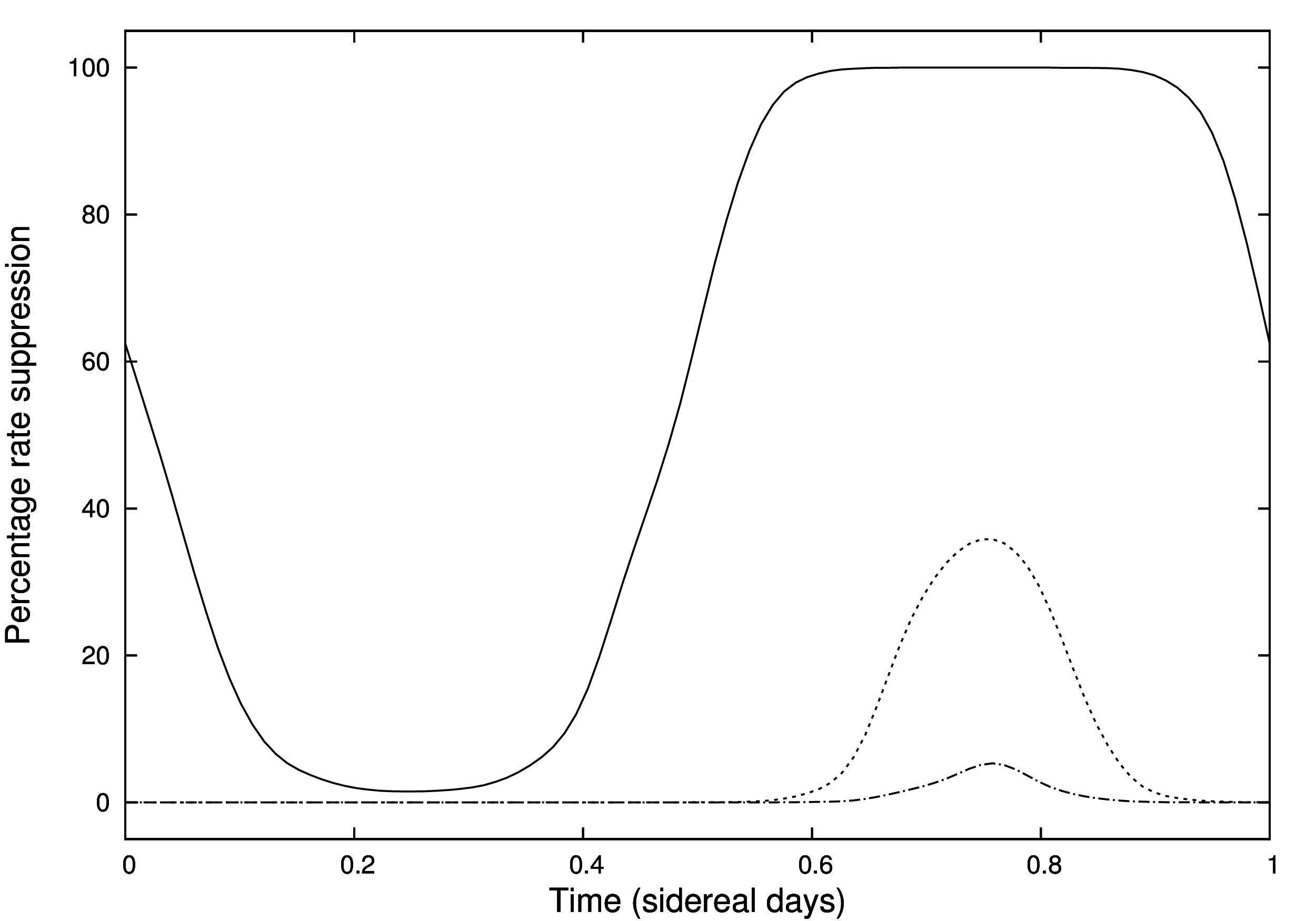, angle=270,width=14.5cm}}
\vskip -0.9cm
\noindent
{\small
Figure 3: Percentage rate suppression for a detector situated in the Stawell mine 
for $m _{F_2} = 10 \ \rm GeV$ (solid line), $m _{F_2} = 100 \ \rm GeV$ (dashed line), $m _{F_2} = 1 \ \rm TeV$ (dot-dashed line). We have assumed $\alpha ' = 10 ^{-2}$, $|Z'| = 10$, a recoil energy of 2 keV and an Na target ($m _T \simeq 23 m _p$).}

\indent
\vskip 0.3cm
\centerline{\epsfig{file=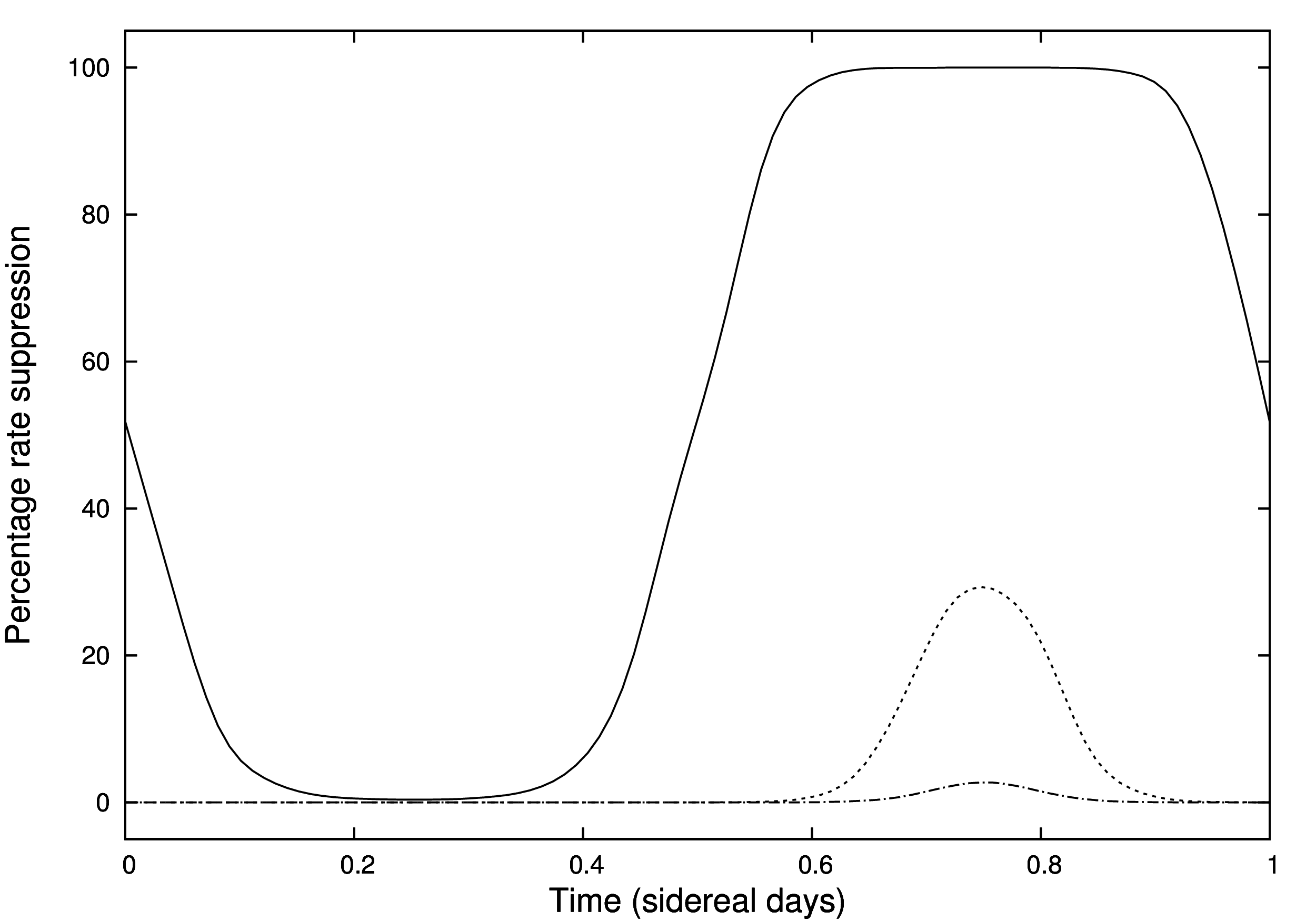, angle=270,width=14.5cm}}
\vskip -0.9cm
\noindent
{\small
Figure 4: Same parameters as per Figure 3 except for a detector located in the Andes Lab.}
\vskip 0.4cm
%
An important quantity is ${\cal R} _{\max}$, the maximum value the percentage rate suppression reaches during the course of a sidereal day. In principle, ${\cal R} _{\max}$ is expected to depend on six parameters. Three of them are fundamental: $m _{F_2}$, $\alpha '$ and $Z'$. The other three are instead related to the experimental setup: $m _T$, $E _R$ and $\theta _l$, that is, the mass of the target nuclei, the relevant recoil energy and the latitude of the detector. To simplify our analysis, we shall fix $\theta _l$ focusing on two latitudes of particular interest: $\theta _l \simeq 37.1 ^{\circ}$ and $\theta _l \simeq 30.2 ^{\circ}$, as discussed above. Further, our numerical analysis determines that ${\cal R} _{\max}$ manifests a very minor dependence on the target nuclei mass, $m _T$, and the recoil energy, $E _R$, for recoil energies in the range $0.1 \ {\rm keV} \lesssim E _R \lesssim 20 \ {\rm keV}$. The end result is that, at a fixed latitude, ${\cal R} _{\max}$ depends mainly on the three fundamental parameters: $m _{F_2}$, $\alpha '$ and $Z'$. For detectors located in the Stawell mine and at the Andes Lab, we find that the maximum percentage suppression rate during the course of a sidereal day can be roughly approximated by:
\begin{eqnarray}
{\cal R} _{\max} & \approx & \min \left [ 55\left ( \frac{\alpha '}{10 ^{-3}} \right ) ^{0.1}\left ( \frac{m _{F_2}}{50 \ \rm GeV} \right ) ^{-0.9}\left ( \frac{|Z'|}{10} \right ) ^{0.6} \% \ , \ 100 \% \right ] \ ({\rm Stawell})  \ , \nonumber \\
{\cal R} _{\max} & \approx & \min \left [ 40\left ( \frac{\alpha '}{10 ^{-3}} \right ) ^{0.1}\left ( \frac{m _{F_2}}{50 \ \rm GeV} \right ) ^{-0.9}\left ( \frac{|Z'|}{10} \right ) ^{0.6} \% \ , \ 100 \% \right ] \ ({\rm Andes}) \ .
\label{rmax}
\end{eqnarray}
The above results hold approximately within the region of parameter space: $5 \times 10 ^{-4} \lesssim \alpha ' \lesssim 5 \times 10 ^{-2}$, $5 \ {\rm GeV} \lesssim m _{F_2} \lesssim 300 \ {\rm GeV}$, $5 \lesssim |Z'| \lesssim 40$. Observe that for both of these locations ${\cal R} _{\max} \gtrsim 10\%$ for nearly all of this parameter space.


\section{Conclusion}

Dissipative hidden sector dark matter appears to be a viable and interesting scenario which has the potential to explain the observed properties of galaxies (as well as large-scale structure). This explanation entails nontrivial galactic dynamics with halo dissipative cooling balanced by heating. It has been shown that ordinary core-collapse supernovae can supply the required heating provided that kinetic mixing interaction with strength $\epsilon \sim 10 ^{-9}$ exists. Such kinetically mixed dark matter can be probed by direct detection experiments. The self-interactions imply that this type of dark matter can be captured within the Earth and shield a dark matter detector from part of the halo dark matter wind. We have shown that, because the direction of this wind changes during the day, so does the amount of shielding, thereby giving rise to a diurnal modulation effect. This effect is expected to be particularly enhanced for a detector located in the Southern hemisphere because, for part of the day, the halo dark matter wind travels through the core of the Earth to reach the detector.

We have estimated the size of this effect, by computing the maximum rate suppression due to dark matter capture for two detectors located in the Southern hemisphere [Eqs.(\ref{rmax})]. Interestingly, we have found that for a large range of parameters the maximum percentage rate suppression during the course of a sidereal day can be large ($\gtrsim 10\%$). Such an effect can potentially be observed in direct detection experiments located in the Southern hemisphere, and would be a smoking gun for self-interacting dark matter.


\vskip 1 cm


\begin{flushleft}

{\Large \bf Acknowledgments}

\vskip 0.3cm
\noindent
SV would like to thank Jackson Clarke and Brian Le for valuable help with the computational side of this work. This work was partly supported by the Australian Research Council and the Melbourne Graduate School of Science.

\end{flushleft}

\vskip 0.2cm
\noindent


\end{document}